\documentclass{moriond}

\usepackage{graphicx}
\usepackage{epsfig}
\usepackage{bm}
\usepackage{bbold}
\usepackage{pgf}
\usepackage{color}
\usepackage[T1]{fontenc}
\usepackage[latin1]{inputenc}
\usepackage{graphicx}
\usepackage[english]{babel}
\usepackage{amsmath}
\usepackage{amssymb}
\usepackage{amsfonts}
\usepackage{makecell}
\usepackage{hyperref}
\usepackage[draft,deletedmarkup=xout]{changes}
\usepackage{mathtools}
\usepackage{soul}
\usepackage{ulem}
\usepackage{relsize}

\def\be{\begin{equation}}
\def\ee{\end{equation}}
\def\bea{\begin{eqnarray}}
\def\eea{\end{eqnarray}}

\newcommand{\Photo}{}

\begin{document}
\vspace*{4cm}
\title{The primordial structure from Quantum Cosmological bouncing models}

\author{ Jaime de Cabo Martin }

\address{National Centre for Nuclear Research, Pasteura 7, 02-093
Warszawa, Poland}
\maketitle

\abstracts{
By quantizing the background as well as the perturbations in a simple one fluid cosmological model, we show that there exists an ambiguity in the choice of relevant variables, potentially leading to incompatible observational physical predictions. In a classical inflationary background, the exact same canonical transformations lead to unique predictions, so the ambiguity we put forward demands a semiclassical background with a sufficiently strong departure from classical evolution. The latter condition is clearly satisfied by bouncing models. We propose coherent states as the tool for introducing the semiclassical universe. We solve the quantum dynamics of the perturbation modes both analytically and numerically and investigate the amplitude spectra of the perturbations. We study the underlying quantum state, the Bunch-Davies vacuum, from the point of view of late-time observers by means of the Bogolyubov transformations. In particular, we study the phase space probability distributions obtained with the standard coherent states built from instantaneous vacua. We discuss the issue of the temporal phase shift with which the modes emerge from the bounce as sine waves. Finally, we find that the model may be fitted to data and shed light on the physical universe, constraining free parameters of the bouncing universe.}

\section{Introduction}
Observational data from the Cosmic Microwave Background indicates that the Universe has emerged from its primordial phase furnished with small adiabatic density perturbations with a nearly scale-invariant amplitude spectrum, providing  a direct proof on the large scale (or small mode) power suppression. The most accepted theoretical explanation to the current observations is the idea that they seem to be coming from quantum vacuum fluctuations of single scalar field added to Einstein-Hilbert action, and this is compatible with inflation scenario, that supposes a phase of exponential accelerating expansion with constant Hubble rate. Moreover, the inflationary paradigm help us to solve some issues of the Big Bang theory, for instance the horizon or the flatness problem, homogeneity, isotropy, etc. However, there exist some problems that are not addressed by inflation, specially the so-called singularity problem, and here is where the motivation for studying the bouncing cosmological solutions arises. We propose the construction a theory of the primordial universe based on the assumption that it
was dominated by quantum gravity effects, which led the Universe to avoid the initial singularity when quantizing the model. This would create
a physical mechanism to generate the primordial structures and started the cosmological expansion. In the future, these description will also allow us to construct a theory that will need a smaller number of primordial symmetries, with new features like anisotropic evolution.
The results presented here represent a summary of the present work in d. C. Martin \textit{et al}.\cite{ja}, and an extension included in the forthcoming work \textit{Quantum bounce models and the primordial structure}, along with P. Ma\l kiewicz (NCBJ, Warsaw) and P. Peter (IAP, Paris).

The basis for obtaining a quantum cosmological theory is to write  General Relativity in Hamiltonian formulation. It was proven\cite{PM} that, starting from the usual 3+1 foliation of the GR ADM-metric: $ds^2=-N^2dt^2+q_{ab}(dx^a+N^adt)(dx^b+N^bdt)$, the scalar physical Hamiltonian for a FLRW universe filled with one fluid, up to second order in perturbations, is found to read:
\begin{equation}
    {\textbf{H}=\int N\mathcal{H}_0\big|^{(0)}+\sum_k
N\left(\kappa_0\frac{w(w+1)ap^2}{6}\Pi^2_k+\frac{3}{2(w+1)}\frac{a^{-3}k^2\mathcal{V}_0}{\kappa p^2}\Phi^2_k\right)}
\end{equation}
with $\kappa_0=8\pi G_N/\mathcal{V}_0$, $w$ representing the kind of fluid, and $a$, $p$, our the canonical variables. It consist in a zeroth-order background which is a first class constraint, and a second order part in terms of the Dirac observables $\Phi$, $\Pi$ that are the gauge-invariant scalar perturbation variables.

\section{Obtaining the quantum theory}
\subsection{Classical Solution}
Our background variables can be redefined into new variables $q$, and $p$ in terms of the fluid Friedmann time $\tau$ in order to simplify the Hamiltonian and equations of motion. The perturbation variables can also be redefine in many (classically equivalent) different ways, each one yielding to a different parametrization of the Hamiltonian. In our work we just considered two of them: Mukhanov-Sasaki (or \textit{Conformal}) param. with $v_k=Z\phi_k$, and Natural (or \textit{Fluid}) param. with $\phi_k=\Phi_k/(\mathcal{V}_0\sqrt{Z})$. We can also conveniently redefine our time variable into a conformal time $\eta$, so that we can write the equations of motion in the usual way of the Mukhanov-Sasaki equation $v_k''+(wk^2-(q^r)''/q^r)v_k=0$ where there are two classically equivalent choices of $r$: $r_1=(3w-1)/(3-3w)$ and $r_2=2/(3-3w)$. The background solutions are just straight lines with constant momentum in the half-plane phase space that terminate or emerge from singularity. The solution for the perturbations is finite and physically equivalent for both parametrizations but discontinuous at the singularity $\eta=0$
\subsection{Quantization and Semiclassical portrait}
We want to quantize both perturbations and background spacetime. For the perturbations we perform the usual HO quantization process in the same way that is made for inflationary models. For the background we use a formalism called \textit{affine coherent state quantization}\cite{JP}, which, when applied to the selected observables $q$ and $p$, makes an extra repulsive term naturally arise, resolving the singularity and being the term responsible for the generation of bouncing trajectories in the phase space (Fig. 1 in de Cabo Martin \textit{et al}.\cite{ja}).

After that, assuming  the full state vector to be a product of background and perturbation
states without backreaction, we produce a class of semiclassical (or semi-quantum) description for the background\cite{AM}. We just don't look for a generic solution of the background but in the space of coherent states, computing the expectation value of the Hamiltonian inside them, and generating the dynamics. We end up with two different semiclassical Hamiltonians for each parametrization, and we will solve the perturbations quantum dynamics  within this portrait. 

Thus, when we solve the Heisenberg equations of motion of perturbations for each of both parametrizations, we obtain two Mukhanov-Sasaki equations with two different potential terms  $\mathcal{V}=(q^r), _{\eta\eta}/q^r$, for natural param. being $r=r_1$ and for M-S being $r=r_2$. Both potential have different shape in time close to the bounce and coincide in the classical limit $|\eta|\rightarrow\infty$, meaning we obtain two inequivalent quantum theories with different predictions for every mode $k_{N/MS}$ (see Fig. 3 and Fig. 4 in de Cabo Martin \textit{et al}.\cite{ja}).

\section{Amplitude spectrum of curvature perturbations}
\begin{figure}[h]
\centering
\includegraphics[scale=0.15]{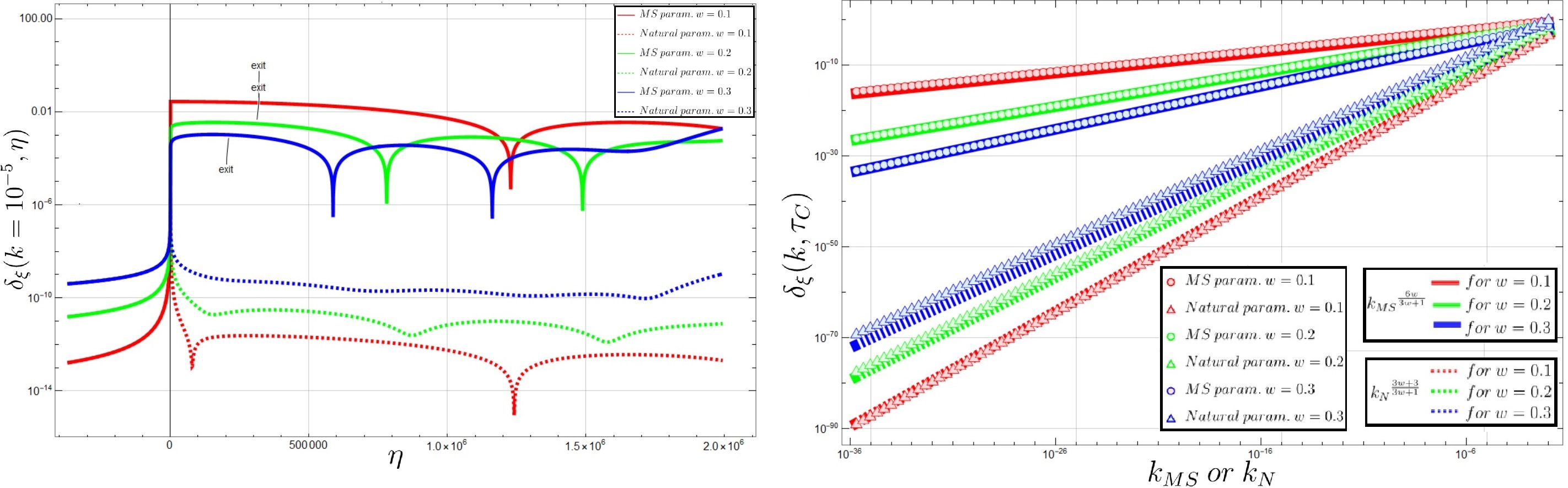}
\caption{\textit{Left}: amplitude spectrum of
  perturbations evolution in $\eta$ for a single mode with
  $k_{N/MS} =10^{-5}$, for three different fluids $w$. \textit{Right}: (in logarithmic scale) numerical and analytical primordial power
  spectrum vs. $k_{N/MS}$ for both parametrizations and different kind of fluid $w$.}
\label{fig:amplitude_ev} 
\end{figure}

We want to solve the quantum dynamics of the perturbations, and we do it both numerically and analytically, in order to compute the amplitude spectrum of the perturbations defined as: $\delta_{\xi}[k,\tau(\eta)]=\hbar\sqrt{\mathcal{V}_0}\frac{|v_{|k|}|}{2\pi a}k^{3/2}$. In order to do that we need some expressions for the initial conditions coming from the minimisation of the Hamiltonian. The process is similar for both parametrizations. First we do it numerically, and we obtain its value for a wide discrete set of modes just after the moment they exit the potential, because we expect its value to get frozen during its evolution (Fig. \ref{fig:amplitude_ev} left) for a significant fraction of its period after that moment. That value is what we shall call primordial spectrum, and when we compute it in terms of the different modes (Fig. \ref{fig:amplitude_ev} right) we clearly obtain a different sets of slopes depending on $w$ for each parametrization, representing two different values of the spectral index: $n_S=6w/(3w+1)$ for M-S param. and $n_S=(3w+3)/(3w+1)$ for natural. 

For the analytical integration we need to take some approximation in order to solve the complex equations, the one being that we are interested only in very small modes that crossed the potential far before the bounce when it could be assumed as in its classical limit. Then, we need match those solutions with the known vacuum ones at the initial crossing potential time. We compute them at the exit crossing time to obtain the formulas for the primordial spectrum:
\begin{eqnarray}
  &&\delta_{\xi}^{MS}=\frac{\sqrt{\mathcal{V}_0\mathfrak{M}_\textsc{s}}\omega f}{2 
  w^\frac{3}{2}}
  \left(\frac{q_\textsc{b}\omega f}{\gamma}\right)^{\frac{-6w}{1+3w}}
  \left|\frac{2}{3-3w}C+(w+1)fD\right|k_{MS}^{\frac{6w}{3w+1}}\label{analytic_spectrum_Phi}\\
  &&\delta_{\xi}^N=\frac{\sqrt{\mathcal{V}_0}}{2\pi w^{\frac{3}{2}}}
  \left(\frac{q_\textsc{b}\omega f}{\gamma}\right)^{\frac{-2}{1+3w}}
  \left|\frac{3w+1}{3-3w}C+2(w+1)fD\right|k_{N}^{\frac{3w+3}{3w+1}}\label{analytic_spectrum_v}
\end{eqnarray}
where $C, D$ are normalization constants, $f=\sqrt{2(1-3w)}/3(1-w^2)$ and $q_B, \omega$, $\gamma$ and $\mathfrak{M}_\textsc{s}$ are defined in [1]\cite{ja}. We obtain the same $n_s$ as from the numerical calculations. It is worthy to mention that in M-S param. we found the same spectral index that for primordial gravitational waves\cite{AM}.
\section{Final quantum state}
\subsection{Particle distributions}
In what follows we study further differences and similarities between the respective final states. For a fixed mode, we characterize using different approaches. We start using particle distributions. By relating different mode expansions of the perturbation fields $v(x,\eta)$ with different creation and annihilation operators $\hat{a}^\dagger$, $\hat{a}$ and $\hat{b}^\dagger$, $\hat{b}$ we can introduce the known Bogolyubov coefficients $\alpha$ and $\beta$.
Via the Bogolyubov transformations we can express the initial Bunch Davies vacuum of $a$-particles with respect to the $b$-particles instantaneous vacuum state that an observer will see at a later time.
Hence, although we start with a state of no $a$-particles, this same state ends up containing $b$-particles with quantas of left and right moving waves forming a standing wave, yielding to particle production. 
We can compute the occupation number of the $n$-particle state of that wave, 
as a function of $n$ representing the probability distribution of particles at the exit crossing time.
\subsection{Phase space representation and temporal phase shift}
\begin{center}
    \begin{figure}[h]
        \centering
       \includegraphics[scale=0.26]{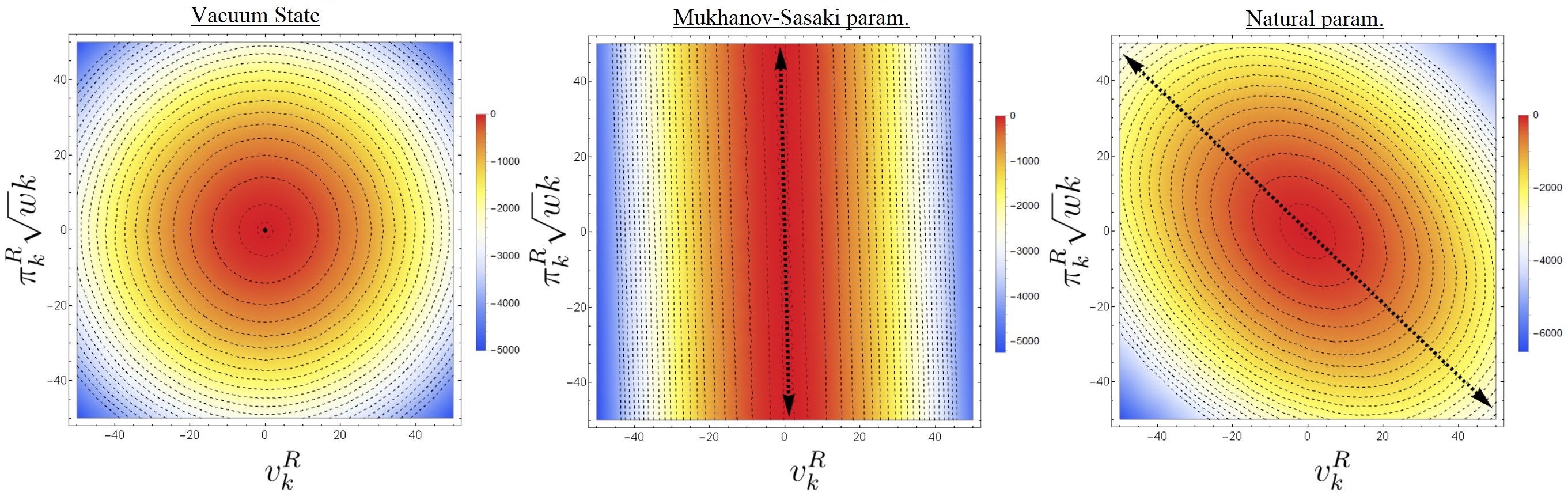}
        \caption{Phase space probability distribution
        for $w=0.28$ and
        $k=10^{-5}$. \textit{Left:} The initial vacuum state. \textit{Center:} M-S param. squeezed state at exit crossing time $\eta_c$. \textit{Right:} Natural param. squeezed state at $\eta_c$.}   
      \label{fig:ellipses} 
    \end{figure}
\end{center}

The next analysis would begin by decomposing the Fourier components of the field in real and imaginary part, since they describe two modes of the standing wave. The final state of both modes can be expressed in the standard coherent state representation by action of a this displacement operator  on $b$-vacuum. We compute the probability distribution for the final state starting from fixed Bunch-Davies vacuum which underlies evolution in Heisenberg picture (Fig. \ref{fig:ellipses}). Once the system evolves through the bounce, the vacuum gets excited and particles are produced leading to final squeezed states. The most important result is that the value of the primordial amplitude is of quantum nature, different for each parametrization and has non-zero uncertainty, but there is a temporal phase of oscillation with which this amplitude emerges when exiting the potential which is classical and represented in Fig. \ref{fig:ellipses} by the black arrows.
This phase shift can be obtained analytically, and it plays the role of making the perturbation fields behave as sine waves when the gravitational potential vanish and radiation era begins, what is a very important feature to ensure the constancy of primordial amplitude of the comoving curvature $\delta R_k\propto \sin(\sqrt{w}k\eta)/\eta\approx\sqrt{w}k$, as inflation does. It can be also proven that this phase shift is scale independent for large scales so that all modes emerge more or less coherently.
\section{Conclusions}
We are able to obtain models of the primordial universe avoiding the singularity problem.
Also, we found that classically equivalent parametrizations lead to two unitarily inequivalent quantum theories. This dependence on parametrizations is a natural consequence of the non-linearity of the theory of gravity, which yields to this ambiguity when we perform quantization.
Moreover, we found that these models may be fitted to data of the primordial spectrum, and are able to explain the constancy of curvature perturbations like inflationary models. Finally, writing the results \eqref{analytic_spectrum_Phi}, \eqref{analytic_spectrum_v} in terms of physical parameters  $\delta_{\xi}^{MS}(k_*)\propto r/{\sqrt{\mathfrak{K}}}$, the data from Planck observations allow us to set constraints on features like the strength of the bounce $\mathfrak{K}$ or size ratio between entire-observable universe $r$, depending on the kind of fluid $w$ governing the dynamics.

\end{document}